\newif\ifarxiv
\arxivtrue

\documentclass[10pt,conference]{IEEEtran}
\IEEEoverridecommandlockouts

\usepackage[T1]{fontenc}
\usepackage[utf8]{inputenc}
\usepackage{bm}
\usepackage{cite}
\usepackage{booktabs}
\usepackage{amsmath,amssymb,amsfonts}
\usepackage{mathtools}
\usepackage{graphicx}
\usepackage{textcomp}
\usepackage[table]{xcolor}
\usepackage{microtype}
\usepackage{enumitem}
\usepackage[breaklinks=true,hidelinks,hyperfootnotes=false]{hyperref}
\usepackage[nameinlink,capitalize]{cleveref}

\usepackage{tikz}
\usetikzlibrary{calc,arrows.meta,positioning}

\tikzset{
 block/.style={draw, rounded corners=2.5pt, very thick, fill=white,
  align=center, inner sep=3pt},
 sblock/.style={block, minimum width=1.45cm, minimum height=0.92cm},
 head/.style={draw, rounded corners=2pt, thick, fill=white,
  minimum width=2.05cm, minimum height=0.66cm, align=center,
  font=\scriptsize, inner sep=2pt},
 fusion/.style={block, minimum width=3.25cm, minimum height=1.92cm,
  text width=3.10cm},
 fast/.style={block, minimum width=3.30cm, minimum height=1.15cm,
  text width=3.10cm},
 panel/.style={draw, rounded corners=3pt, thick, fill=white,
  minimum width=4.10cm, minimum height=2.30cm, align=center},
 circ/.style={draw, circle, thick, fill=white, inner sep=0pt},
 arr/.style={-{Latex[length=2.15mm,width=1.6mm]}, thick, shorten >=2.3pt, shorten <=2.0pt},
 darr/.style={-{Latex[length=2.15mm,width=1.6mm]}, thick, dashed, shorten >=2.3pt, shorten <=2.0pt},
 lab/.style={font=\scriptsize, align=center},
 legendlab/.style={font=\footnotesize, align=center},
 titlelab/.style={font=\bfseries\small, align=center}
}

\newcommand{\daruancell}[2]{%
 \begin{scope}[shift={(#1,#2)}]
  \draw[thick] (-0.145,-0.145) rectangle (0.145,0.145);
  \draw[thick,-{Latex[length=1.15mm,width=0.85mm]}] (-0.112,-0.112) -- (0.115,0.115);
  \draw[thick] (0,0) circle (0.070);
 \end{scope}%
}

\usepackage{orcidlink}

\def\BibTeX{{\rm B\kern-.05em{\sc i\kern-.025em b}\kern-.08em
  T\kern-.1667em\lower.7ex\hbox{E}\kern-.125emX}}

\ifarxiv
\renewcommand{\baselinestretch}{1.02}
\fi

\setlist[itemize]{leftmargin=*,nosep}

\begin{document}

\title{Complementary Matrix-Gated QKAN Fast-Weight Programmers for Quantum Dynamics Forecasting
\ifarxiv
\thanks{The views expressed in this article are those of the authors and do not represent the views of Wells Fargo. This article is for informational purposes only. Nothing contained in this article should be construed as investment advice. Wells Fargo makes no express or implied warranties and expressly disclaims all legal, tax, and accounting implications related to this article.}
\fi
}

\ifarxiv

\author{
\IEEEauthorblockN{
  Kuo-Chung Peng$^{1,2,}$\IEEEauthorrefmark{1}\orcidlink{0009-0001-8342-2481},
  Samuel Yen-Chi Chen$^{3,}$\IEEEauthorrefmark{1}\orcidlink{0000-0003-0114-4826},
  Jiun-Cheng Jiang$^{1,4,5}$\orcidlink{0009-0005-1134-4962},
  Chen-Yu Liu$^{6}$\orcidlink{0000-0002-5437-5188},
  En-Jui Kuo$^{7}$\orcidlink{0000-0002-6770-0285}, \\
  Yun-Yuan Wang$^{4}$\orcidlink{0009-0001-0323-3382}, 
  Tzung-Chi Huang$^{4}$,
  Prayag Tiwari$^{8}$\orcidlink{0000-0002-2851-4260},
  Chi-Sheng Chen$^{9}$\orcidlink{0000-0003-0807-0217},
  Chun-Hua Lin$^{1,2}$\orcidlink{0009-0002-4383-0453},
  Yu-Chao Hsu$^{2,10}$\orcidlink{0009-0004-7221-3854}, \\ 
  Tai-Yue Li$^{2}$\orcidlink{0000-0002-1993-1863}, 
  Saif Al-Kuwari$^{12}$\orcidlink{0000-0002-4402-7710}, 
  Simon See$^{11}$\orcidlink{0000-0002-4958-9237},
  Kuan-Cheng Chen$^{12,}$\IEEEauthorrefmark{2}\orcidlink{0000-0002-6575-7034},
  Nan-Yow Chen$^{2,}$\IEEEauthorrefmark{3}\orcidlink{0000-0001-8139-6809},
  Hsi-Sheng Goan$^{1,5,6,13,}$\IEEEauthorrefmark{4}\orcidlink{0000-0001-8117-5846}
}
\IEEEauthorblockA{$^1$Department of Physics and Center for Theoretical Physics, National Taiwan University, Taipei, Taiwan}
\IEEEauthorblockA{$^2$National Center for High-Performance Computing, National Institutes of Applied Research, Hsinchu, Taiwan}
\IEEEauthorblockA{$^3$Wells Fargo, New York, NY, USA}
\IEEEauthorblockA{$^4$NVIDIA AI Technology Center, NVIDIA Corp., Taipei, Taiwan}
\IEEEauthorblockA{$^5$Center for Quantum Science and Engineering, National Taiwan University, Taipei, Taiwan}
\IEEEauthorblockA{$^6$Graduate Institute of Applied Physics, National Taiwan University, Taipei, Taiwan}
\IEEEauthorblockA{$^7$Department of Electrophysics, National Yang Ming Chiao Tung University, Hsinchu, Taiwan}
\IEEEauthorblockA{$^8$School of Information Technology, Halmstad University, Sweden}
\IEEEauthorblockA{$^{9}$Beth Israel Deaconess Medical Center \& Harvard Medical School, Boston, MA, USA}
\IEEEauthorblockA{$^{10}$School of Electrical Engineering, Korea Advanced Institute of Science and Technology, Daejeon, Korea}
\IEEEauthorblockA{$^{11}$NVIDIA AI Technology Center, NVIDIA Corp., Singapore, Singapore}
\IEEEauthorblockA{$^{12}$Qatar Center for Quantum Computing, College of Science and Engineering, Hamad Bin Khalifa University, Doha, Qatar}
\IEEEauthorblockA{$^{13}$Physics Division, National Center for Theoretical Sciences, Taipei, Taiwan}

\IEEEauthorblockA{\IEEEauthorrefmark{1}These authors contributed equally to this work.}
\IEEEauthorblockA{Emails: \IEEEauthorrefmark{2}\href{mailto:kchen@hbku.edu.qa}{kchen@hbku.edu.qa}, 
\IEEEauthorrefmark{3}\href{mailto:nanyow@nchc.narl.org.tw}{nanyow@nchc.narl.org.tw},
\IEEEauthorrefmark{4}\href{mailto:goan@phys.ntu.edu.tw}{goan@phys.ntu.edu.tw}
}
\vspace{-30pt}
}
\else
\author{Anonymous Authors}
\fi

\maketitle

\begin{abstract}

Sequence models must decide what to write into memory and what to retain. In quantum and quantum-inspired sequence learning, nonlinear recurrent updates often require repeated circuit evaluations and sequential backpropagation through time, making long contexts costly. Gated fast-weight programmers (FWPs) based on quantum-inspired Kolmogorov–Arnold networks (QKANs) alleviate this bottleneck by storing context in time-varying fast parameters. However, their scalar gate applies one retention–write balance to every fast-state coordinate, forcing all parameters to share a memory timescale. We introduce Self-Modulating QKAN-based FWPs, which replace this broadcast gate with low-rank-generated element-wise modulation of the new-proposal branch, a bounded old-state branch, or both. We further propose Complementary Matrix Gating (CMG), which uses one sigmoid matrix gate to retain the old state and its complement to write the new proposal. CMG provides coordinate-wise memory control while preserving the bounded convex update and affine prefix-scan structure of scalar gating, at the modulation-head cost of a single-branch rule. We compare four self-modulating rules with scalar gating across four FWP architectures combining classical and QKAN-based slow and fast programmers. Across seven single-step forecasting benchmarks and five sequence lengths, CMG gives the most consistent improvements for architectures whose fast programmer incorporates a QKAN-based module. In direct multi-step forecasting of Jaynes–Cummings and transmon–resonator dynamics simulated with CUDA-Q Dynamics, CMG models maintain mean-squared errors on the order of 0.001 or lower across forecasting horizons of 4, 8, and 16 steps, while improving on their scalar-gated counterparts by at least 91.2\%. These results establish coordinate-wise complementary modulation as a stable and effective update for QKAN-based FWPs.

\end{abstract}


\vspace{-10pt}
\section{Introduction}
\label{sec:introduction}

Sequence models must decide what to write into memory and what to retain from the past. 
Long short-term memory (LSTM) networks and gated recurrent units (GRUs) make this trade-off explicit with input, update, and forget gates \cite{hochreiter1997lstm,gers2000learning,cho2014learning,chung2015gated}. 
In quantum machine learning (QML), recurrent memory can be especially expensive: nonlinear recurrent updates generally require unparallelizable, step-by-step circuit evolution, rendering backpropagation through time (BPTT) over long sequences highly time-consuming~\cite{bausch2020recurrent,chen2022qlstm,takaki2021learning,pascanu2013difficulty,schuld2019evaluating}.
Fast-weight programming offers a more parallelizable alternative by storing temporal context in dynamically updated parameters rather than in a recurrent hidden state \cite{schmidhuber1992fastweight,ba2016fastweights,schlag2021linear}. Quantum Fast Weight Programmers (QFWPs) implement this idea with a slow classical programmer that updates a fast variational quantum circuit (VQC) \cite{chen2024qfwp}. 
Quantum-inspired Kolmogorov--Arnold Network (QKAN)-based Fast-Weight Programmers (QKAN-FWPs) improve scalability by replacing multi-qubit fast circuits with QKAN modules, whose DatA Re-Uploading ActivatioN (DARUAN) edge functions are single-qubit data re-uploading circuits \cite{peng2026gatedqkanfwp,jiang2025qkan,liu2025kan,perez2020data}.
While in ref.~\cite{peng2026gatedqkanfwp}, scalar gates are stable and parameter-efficient, broadcasting a single retention/write coefficient to every fast-state coordinate, forcing all fast parameters to share the same memory timescale. 
The central question of this paper is therefore whether QKAN-FWPs can benefit from coordinate-wise memory control without losing boundedness or the affine parallel-prefix structure that makes fast-weight recurrences efficient \cite{blelloch1990prefix}.

We answer this question by implementing self-modulating fast-state updates in QKAN-FWP and by introducing Complementary Matrix Gating (CMG) as a new self-modulating rule. 
Following the self-modulating QFWP~\cite{chen2026selfmod}, we let the slow programmer emit low-rank-generated element-wise modulators for the new-update branch, the old-state branch, or both, yielding Only-new, Only-old, and Full variants. 
We bound the old-state modulation branch with $\tanh$ so each coordinate can retain, suppress, or sign-adjust memory without geometric amplification~\cite{peng2026smqfwp_bounded}. 
CMG, in contrast, ties old-state retention and new-state writing through one sigmoid matrix gate. 
The gate weights the old fast state, and its complement weights the new proposal. 
Thus CMG is an element-wise generalization of gated QKAN-FWPs that preserves its bounded convex update and affine scan-compatible form. 
Empirically, CMG is the most reliable update rule for single-step prediction benchmarks. 
Across seven benchmarks and five sequence lengths, it gives the most consistent improvements over scalar gating.
On Jaynes--Cummings and transmon--resonator quantum dynamics direct multi-step forecasting \cite{kim2023cuda}, models that deploy the CMG update rule maintain prediction mean-squared error(MSE) of order $10^{-4}$ or lower across horizons, improving over gated update rule by at least 91.2\%.
Only-old and Full are competitive but either lack complementary write control or require two modulation matrices.

In summary, we introduce CMG, a coordinate-wise self-modulating update rule for QKAN-FWPs that preserves bounded convex dynamics and prefix-scan compatibility. 
Through systematic comparisons with write-side, memory-side, and full self-modulation, we show that coordinate-wise memory control yields more reliable gains than scalar gating, with CMG providing the most consistent improvements across single-step and direct multi-step forecasting benchmarks.

\section{Related Work}

\paragraph{QML sequence models and fast-weight memory}
QML sequence models commonly insert quantum reservoirs, recurrent quantum circuits, or VQCs into classical temporal architectures for time-series forecasting or reinforcement learning tasks~\cite{fujii2017harnessing,bausch2020recurrent,chen2022qlstm,takaki2021learning,li2023qrnn,li2024qgrnn, kundu2026qrl}. 
These models can be expressive, but long sequence lengths still require repeated temporal circuit evaluation. 
Fast-weight programming instead accumulates context in adaptive fast parameters \cite{schmidhuber1992fastweight,ba2016fastweights,schlag2021linear}. 
Self-modulating QFWPs further extend QFWP with element-wise fast-parameter gating~\cite{chen2024qfwp,chen2026selfmod,peng2026smqfwp_bounded}.

\paragraph{KAN/QKAN sequence models and element-wise gating}
Kolmogorov--Arnold networks (KANs) replace fixed node activations and linear weights with learnable univariate edge functions, inspiring KAN forecasters, temporal KAN cells, mixture-of-expert variants, and decomposition or frequency modules for time series~\cite{liu2025kan,xu2024kan_ts,han2024kan4tsf,huang2025timekan,Amir2026kanreview}.
QKAN substitutes spline-style edge functions with DARUAN functions, combining KAN edge-wise function learning with parameter-efficient quantum-inspired nonlinear modeling~\cite{perez2020data,jiang2025qkan}. 
Recent QKAN sequence models incorporate QKAN blocks into LSTM-style gates, fast-weight programmers, or transformers~\cite{hsu2026qkanlstm,peng2026gatedqkanfwp,peng2026qkanfwp_TM,jiang2025qkan,lin2026gqkae}. 
Coordinate-wise gating is well established in recurrent models, with LSTMs and gated recurrent units using element-wise gates to balance retained memory and candidate states~\cite{hochreiter1997lstm,cho2014learning}. 
Later models extend this feature-wise control through lightweight recurrence, adaptive decay, or element-wise attention~\cite{schlag2017gated, lei2018simple,che2018grud,zhang2019eleatt}. 
CMG bridges gated QKAN-FWPs and self-modulating QFWPs through low-rank-generated matrix gates with complementary old/new weighting.

\section{Methods}
\label{sec:methods}

\subsection{QKAN and HQKAN programmer backbones}
Fast-weight programming separates a slow programmer from a fast programmer. 
Let $x_t$ be the scalar input, $S_\psi$ the slow programmer, and $F(\cdot;\Theta_t)$ the fast programmer with time-dependent fast state $\Theta_t$. 
The slow programmer reads $x_t$ and updates fast states, $\Theta_t$, while the fast programmer predicts with the updated state. 
Temporal information is stored in the trajectory of $\{\Theta_t\}$ rather than in a recurrent hidden state.

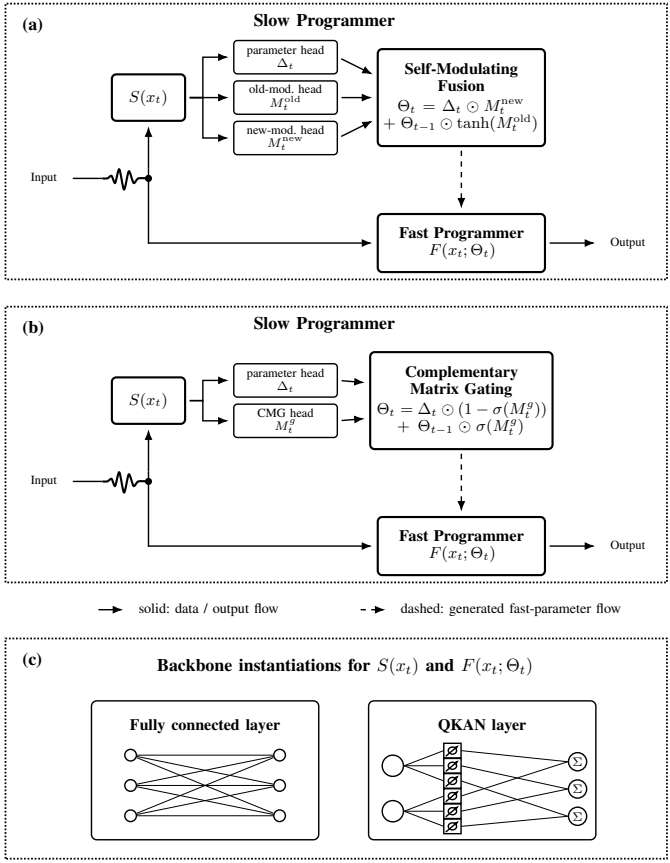
\begin{figure}[t]
  \centering
  \resizebox{\columnwidth}{!}{%
    \begin{tikzpicture}[font=\small, x=1.18cm, y=1cm]

\draw[densely dotted, line width=1.0pt] (-0.35,5.32) rectangle (10.75,10.78);
\draw[densely dotted, line width=1.0pt] (-0.35,-0.68) rectangle (10.75,4.78);
\draw[densely dotted, line width=1.0pt] (-0.35,-6.15) rectangle (10.75,-1.79);

\node[font=\bfseries, anchor=north west] at (-0.17,10.62) {(a)};
\node[font=\bfseries, anchor=north west] at (-0.17,4.62) {(b)};
\node[font=\bfseries, anchor=north west] at (-0.17,-1.95) {(c)};

\node[font=\bfseries] at (5.00,10.43) {Slow Programmer};

\node[lab, anchor=east] at (0.62,7.32) {Input};
\draw[thick] (0.78,7.32) -- (1.28,7.32);
\draw[very thick] plot[smooth] coordinates {
  (1.28,7.32) (1.36,7.32) (1.43,7.37) (1.49,7.23)
  (1.56,7.50) (1.64,7.10) (1.72,7.46) (1.80,7.23)
  (1.88,7.32) (2.05,7.32)};
\fill (2.05,7.32) circle (2.0pt);

\node[sblock] (slowA) at (2.05,8.92) {$S(x_t)$};
\draw[thick] (2.05,7.52) -- (2.05,8.20);
\draw[arr] (2.05,8.20) -- (slowA.south);

\node[head] (deltA) at (4.35,9.72) {parameter head\\[-0.35ex]$\Delta_t$};
\node[head] (moldA) at (4.35,8.92) {old-mod. head\\[-0.15ex]$M_t^{\mathrm{old}}$};
\node[head] (mnewA) at (4.35,8.12) {new-mod. head\\[-0.35ex]$M_t^{\mathrm{new}}$};

\node[fusion] (fusionA) at (7.30,8.92) {\textbf{Self-Modulating}\\[-0.1ex]
\textbf{Fusion}\\[0.25ex]
$\Theta_t=\Delta_t\odot M_t^{\mathrm{new}}$\\[-0.1ex]
$+\;\Theta_{t-1}\odot\tanh(M_t^{\mathrm{old}})$};

\draw[arr] (slowA.east) -- ++(0.28,0) |- (deltA.west);
\draw[arr] (slowA.east) -- (moldA.west);
\draw[arr] (slowA.east) -- ++(0.28,0) |- (mnewA.west);

\draw[arr] (deltA.east) -- ([yshift=0.42cm]fusionA.west);
\draw[arr] (moldA.east) -- (fusionA.west);
\draw[arr] (mnewA.east) -- ([yshift=-0.42cm]fusionA.west);

\node[fast] (fastA) at (7.30,6.04) {\textbf{Fast Programmer}\\[-0.05ex]$F(x_t;\Theta_t)$};
\draw[thick] (2.05,7.52) -- (2.05,6.03);
\draw[arr] (1.98,6.04) -- (fastA.west);
\draw[darr] (fusionA.south) -- (fastA.north);
\draw[arr] (fastA.east) -- (9.62,6.04);
\node[lab, anchor=west] at (9.70,6.04) {Output};

\begin{scope}[shift={(0,-6.00)}]
\node[font=\bfseries] at (5.00,10.43) {Slow Programmer};

\node[lab, anchor=east] at (0.62,7.32) {Input};
\draw[thick] (0.78,7.32) -- (1.28,7.32);
\draw[very thick] plot[smooth] coordinates {
  (1.28,7.32) (1.36,7.32) (1.43,7.37) (1.49,7.23)
  (1.56,7.50) (1.64,7.10) (1.72,7.46) (1.80,7.23)
  (1.88,7.32) (2.05,7.32)};
\fill (2.05,7.32) circle (2.0pt);

\node[sblock] (slowB) at (2.05,8.92) {$S(x_t)$};
\draw[thick] (2.05,7.52) -- (2.05,8.20);
\draw[arr] (2.05,8.20) -- (slowB.south);

\node[head] (deltB) at (4.35,9.32) {parameter head\\[-0.35ex]$\Delta_t$};
\node[head] (mgateB) at (4.35,8.52) {CMG head\\[-0.35ex]$M_t^g$};

\node[fusion, minimum width=3.55cm, text width=3.40cm] (fusionB) at (7.30,8.92) {\textbf{Complementary}\\[-0.1ex]
\textbf{Matrix Gating}\\[0.25ex]
$\Theta_t=\Delta_t\odot(1-\sigma(M_t^g))$\\[-0.1ex]
$+\;\Theta_{t-1}\odot\sigma(M_t^g)$};

\draw[arr] (slowB.east) -- ++(0.28,0) |- (deltB.west);
\draw[arr] (slowB.east) -- ++(0.28,0) |- (mgateB.west);

\draw[arr] (deltB.east) -- ([yshift=0.35cm]fusionB.west);
\draw[arr] (mgateB.east) -- ([yshift=-0.35cm]fusionB.west);

\node[fast] (fastB) at (7.30,6.04) {\textbf{Fast Programmer}\\[-0.05ex]$F(x_t;\Theta_t)$};
\draw[thick] (2.05,7.52) -- (2.05,6.03);
\draw[arr] (1.98,6.04) -- (fastB.west);
\draw[darr] (fusionB.south) -- (fastB.north);
\draw[arr] (fastB.east) -- (9.62,6.04);
\node[lab, anchor=west] at (9.70,6.04) {Output};
\end{scope}

\draw[arr] (1.15,-1.22) -- (1.70,-1.22);
\node[legendlab, anchor=west] at (1.78,-1.22) {solid: data / output flow};
\draw[darr] (5.55,-1.22) -- (6.10,-1.22);
\node[legendlab, anchor=west] at (6.18,-1.22) {dashed: generated fast-parameter flow};

\begin{scope}[shift={(0,-6.35)}]
\node[font=\bfseries] at (5.35,4.00)
  {Backbone instantiations for $S(x_t)$ and $F(x_t;\Theta_t)$};

\node[panel, minimum width=4.5cm, minimum height=2.75cm] (linpanel) at (3.00,1.95) {};
\node[titlelab] at (3.00,2.82) {Fully connected layer};
\foreach \y/\nm in {2.25/li1,1.65/li2,1.05/li3}{\node[circ, minimum size=0.23cm] (\nm) at (1.75,\y) {};}
\foreach \y/\nm in {2.25/lo1,1.65/lo2,1.05/lo3}{\node[circ, minimum size=0.23cm] (\nm) at (4.25,\y) {};}
\foreach \i in {1,2,3}{\foreach \j in {1,2,3}{\draw[thin] (li\i) -- (lo\j);}}

\node[panel, minimum width=4.5cm, minimum height=2.75cm] (hqpanel) at (7.65,1.95) {};
\node[titlelab] at (7.65,2.82) {QKAN layer};
\node[circ, minimum size=0.42cm] (hin1) at (6.15,2.04) {};
\node[circ, minimum size=0.42cm] (hin2) at (6.15,1.14) {};

\foreach \y/\nm in {2.34/cellA,2.04/cellB,1.74/cellC,1.44/cellD,1.14/cellE,0.84/cellF}{
  \coordinate (\nm) at (7.15,\y);
}

\node[circ, minimum size=0.36cm, font=\scriptsize] (sig1) at (9.25,2.12) {$\Sigma$};
\node[circ, minimum size=0.36cm, font=\scriptsize] (sig2) at (9.25,1.58) {$\Sigma$};
\node[circ, minimum size=0.36cm, font=\scriptsize] (sig3) at (9.25,1.04) {$\Sigma$};

\foreach \c in {cellA,cellB,cellC}{\draw[thin] (hin1.east) -- ($ (\c)+(-0.145,0) $);}
\foreach \c in {cellD,cellE,cellF}{\draw[thin] (hin2.east) -- ($ (\c)+(-0.145,0) $);}

\draw[thin] ($ (cellA)+(0.145,0) $) -- (sig1.west);
\draw[thin] ($ (cellC)+(0.145,0) $) -- (sig3.west);
\draw[thin] ($ (cellD)+(0.145,0) $) -- (sig1.west);
\draw[thin] ($ (cellB)+(0.145,0) $) -- (sig2.west);
\draw[thin] ($ (cellF)+(0.145,0) $) -- (sig3.west);
\draw[thin] ($ (cellE)+(0.145,0) $) -- (sig2.west);

\foreach \y in {2.34,2.04,1.74,1.44,1.14,0.84}{\daruancell{7.15}{\y}}
\end{scope}

\end{tikzpicture}%
  }
  \caption{Self-modulating update rules and programmer backbones. (a) Full self-modulation uses separate old- and new-modulation heads. (b) CMG uses a single low-rank-generated matrix gate with complementary weights for the retained and proposed states. (c) The slow and fast programmers may use classical fully connected modules or QKAN-based HQKAN modules. Solid arrows show data/output flow. Dashed arrows show generated fast-parameter flow.}
  \label{fig:self_modulating_qkanfwp}
  \vspace{-15pt}
\end{figure}

We use either classical linear/multi-layer perceptron (MLP) programmers or QKAN-based programmers. 
QKAN replaces spline KAN edge functions with DARUAN functions \cite{jiang2025qkan}. 
For an input $z$, $\phi_\vartheta(z)
  =
  \langle 0|U^\dagger(z;\vartheta)\hat{O}U(z;\vartheta)|0\rangle,$
where $U(z;\vartheta)$ is a parameterized single-qubit data-reuploading unitary and $\hat{O}$ is the measured observable~\cite{perez2020data}. 
In QKAN-based programmers we use a hybrid QKAN (HQKAN) module: a classical encoder maps inputs to a latent space, a QKAN block applies nonlinear transformations, and a classical decoder maps back to the required output dimension. 
Thus the ``QKAN layer'' in \cref{fig:self_modulating_qkanfwp} denotes the quantum-inspired core inside an encoder--QKAN--decoder module. 
\Cref{tab:model_variants} lists the four architectures: fully classical FWP, QKANFWP with an HQKAN fast programmer, QKAN-FWP with an HQKAN slow programmer, and QKAN-QKANFWP with HQKAN in both roles.

\begin{table}[t]
\centering
\caption{Programmer backbones for the four architectural variants. “Classical” denotes a multilayer-perceptron (MLP) slow programmer or a linear fast programmer; “HQKAN” denotes an encoder–QKAN–decoder module.}
\label{tab:model_variants}
\resizebox{\columnwidth}{!}{%
\begin{tabular}{lcc}
\toprule
Model & Slow programmer $S_\psi$ & Fast programmer $F(\cdot;\Theta_t)$ \\
\midrule
FWP & Classical & Classical \\
QKANFWP & Classical & HQKAN \\
QKAN-FWP & HQKAN & Classical \\
QKAN-QKANFWP & HQKAN & HQKAN \\
\bottomrule
\end{tabular}
}
\vspace{-5pt}
\end{table}

\subsection{Gated fast-weight update}
For a generic fast state $\Theta_t$, the slow programmer first proposes $ \Delta_t = S_\psi^{\Delta}(x_t),$
with the same shape as $\Theta_t$. The slow programmer of the gated baseline~\cite{peng2026gatedqkanfwp} also produces $g_t = \sigma(S_\psi^{g}(x_t)), g_t\in[0,1],$
and updates
\begin{equation}
  \Theta_t
  =
  g_t\Theta_{t-1} + (1-g_t)\Delta_t .
  \label{eq:scalar_gated_update}
\end{equation}

\subsection{Low-rank self-modulating updates}
Self-modulation replaces the scalar gate with element-wise modulation. 
For any architecture in \cref{tab:model_variants}, the fast state is reshaped to $\Theta_t\in\mathbb{R}^{P\times Q}$. 
For each branch $r\in\{\mathrm{new},\mathrm{old}\}$, the slow programmer
generates a rank-one modulation matrix from the outer product of two
affine-head outputs:
\begin{equation*}
  M_t^{r}
  =
  m_t^{r,P}\left(m_t^{r,Q}\right)^\top,
  \qquad
  m_t^{r,P}\in\mathbb{R}^{P},
  \quad
  m_t^{r,Q}\in\mathbb{R}^{Q}.
\end{equation*} Following self-modulating QFWP \cite{chen2026selfmod}, the raw update family is
\begin{align*}
  \Theta_t
  &=
  \Delta_t\odot M_t^{\mathrm{new}}
  +
  \Theta_{t-1}\odot M_t^{\mathrm{old}},
  &&\text{Full}, \\
  \Theta_t
  &=
  \Delta_t\odot M_t^{\mathrm{new}}
  +
  \Theta_{t-1},
  &&\text{Only-new}, \\
  \Theta_t
  &=
  \Delta_t
  +
  \Theta_{t-1}\odot M_t^{\mathrm{old}},
  &&\text{Only-old}. 
\end{align*}
The new branch controls write amplitude; the old branch controls coordinate-wise retention, attenuation, amplification, or sign reversal.

Because old-state modulation is multiplied recurrently, we bound it by
\begin{equation*}
  \widetilde{M}_t^{\mathrm{old}}
  =
  \tanh\!\left(M_t^{\mathrm{old}}\right),
  \qquad
  \left|\widetilde{M}_{t,pq}^{\mathrm{old}}\right|\le 1.
\end{equation*}
The bounded Full and Only-old updates are
\begin{align*}
  \Theta_t
  &=
  \Delta_t\odot M_t^{\mathrm{new}}
  +
  \Theta_{t-1}\odot \widetilde{M}_t^{\mathrm{old}},
  &&\text{bounded Full}, \\
  \Theta_t
  &=
  \Delta_t
  +
  \Theta_{t-1}\odot \widetilde{M}_t^{\mathrm{old}},
  &&\text{bounded Only-old}.
\end{align*}
For coordinate $j=(p,q)$, with $a_{t,j}=\widetilde{M}_{t,j}^{\mathrm{old}}$ and $d_{t,j}=[\Delta_t]_j$, bounded Only-old unrolls as
\begin{equation*}
  \theta_{t,j}
  =
  \theta_{0,j}\prod_{u=1}^{t}a_{u,j}
  +
  \sum_{s=1}^{t}
  d_{s,j}
  \prod_{u=s+1}^{t}a_{u,j},
  \qquad
  |a_{u,j}|\le 1.
\end{equation*}
Thus the recurrent kernel cannot amplify stored updates through products of old-state multipliers. The same expression applies to bounded Full after replacing $d_{s,j}$ by $[\Delta_s\odot M_s^{\mathrm{new}}]_j$.

\subsection{Complementary Matrix Gating}
CMG uses the same low-rank factorization but generates one gate matrix,
\begin{equation*}
  M_t^g
  =
  m_t^{g,P}\left(m_t^{g,Q}\right)^\top,
  \qquad
  G_t=\sigma(M_t^g)\in[0,1]^{P\times Q}.
\end{equation*}
The update is
\begin{equation*}
  \Theta_t
  =
  G_t\odot\Theta_{t-1}
  +
  (1-G_t)\odot\Delta_t.
\end{equation*}
Equivalently, CMG replaces the scalar gate in \cref{eq:scalar_gated_update} with an element-wise matrix gate. This is the new self-modulating rule proposed in this paper: the slow programmer generates the gate, but the old and new branches remain complementary. Unlike Full self-modulation, CMG uses one modulation matrix, giving the same modulation-head size as Only-new and Only-old.

CMG inherits the scalar gate's bounded convex update coordinate-wise. For coordinate $j$, let $g_{t,j}\in[0,1]$ and $d_{t,j}=[\Delta_t]_j$. Then
\begin{equation*}
  \theta_{t,j}=g_{t,j}\theta_{t-1,j}+(1-g_{t,j})d_{t,j},
\end{equation*}
and
\begin{equation}
  \theta_{t,j}
  =
  \theta_{0,j}\prod_{u=1}^{t}g_{u,j}
  +
  \sum_{s=1}^{t}
  (1-g_{s,j})d_{s,j}
  \prod_{u=s+1}^{t}g_{u,j} .
  \label{eq:cmg_unroll}
\end{equation}
The nonnegative update weights in \cref{eq:cmg_unroll} sum to $1-\prod_{u=1}^{t}g_{u,j}\le 1$; including the initial-state coefficient gives a convex combination. Hence, under the same bounded-proposal condition used for scalar gating, each coordinate remains within the same bound without requiring a global scalar gate \cite{peng2026gatedqkanfwp}.

All update rules considered here can still be evaluated by parallel prefix scan. After the slow-programmer heads are computed for all time steps, each update has the affine form
\begin{equation*}
  \Theta_t=A_t\odot\Theta_{t-1}+B_t .
\end{equation*}
For Only-new, $A_t=\mathbf{1}$ and $B_t=\Delta_t\odot M_t^{\mathrm{new}}$; for Only-old, $A_t=\widetilde{M}_t^{\mathrm{old}}$ and $B_t=\Delta_t$; for Full, $A_t=\widetilde{M}_t^{\mathrm{old}}$ and $B_t=\Delta_t\odot M_t^{\mathrm{new}}$; for CMG, $A_t=G_t$ and $B_t=(1-G_t)\odot\Delta_t$. The pairs compose associatively,
\begin{equation*}
  (A',B')\circ(A,B)
  =
  (A'\odot A,\; A'\odot B+B'),
\end{equation*}
so the full fast-state trajectory can be obtained with the same affine prefix-scan principle used in fast-weight recurrences~\cite{peng2026gatedqkanfwp}.

\section{Experimental Protocol}
\label{sec:experimental_protocol}

\subsection{Tasks and preprocessing}
We evaluate seven univariate time-series benchmarks and follow the same setup used in prior works~\cite{peng2026gatedqkanfwp,peng2026smqfwp_bounded}. Each scalar sequence is min--max normalized to $[-1,1]$, converted to chronological sliding-window samples, and split into $80\%$ train and $20\%$ test sets.

\paragraph{Damped simple harmonic motion}
The damped simple harmonic motion (SHM) task uses the angular velocity of a nonlinear damped pendulum,
$
  \frac{d^2\theta}{dt^2}
  +
  \frac{b}{m}\frac{d\theta}{dt}
  +
  \frac{g}{L}\sin\theta
  =
  0,
$
with $g=9.81$, $b=0.15$, $L=m=1$, $\theta(0)=0$, and $\dot{\theta}(0)=3$. It tests smooth damped oscillatory prediction.

\paragraph{Bessel function}
The Bessel task uses the second-order Bessel function of the first kind, $J_2(x)$, to test nonlinear approximation with changing amplitude and phase.

\paragraph{NARMA-5 and NARMA-10}
The nonlinear autoregressive moving-average (NARMA) tasks use memory orders $n=5$ and $n=10$ with the standard recurrence
$y_{t+1}
  =
  \alpha y_t
  +
  \beta y_t \sum_{j=0}^{n-1} y_{t-j}
  +
  \gamma u_{t-n+1}u_t
  +
  \delta ,
  \qquad n\in\{5,10\}.$

\paragraph{Delayed quantum control}
The delayed quantum control (DQC) task is a non-Markovian feedback-like signal formed by decaying localized pulses,
$ x(t)
  =
  \sum_{n=0}^{10}
  \exp[-10(t-2n)^2]\exp(-t/16),
  \qquad
  t\in[-2,20].$

\paragraph{Open Jaynes--Cummings dynamics}
This benchmark is an open Jaynes--Cummings system with a two-level qubit coupled to a cavity truncated to 5 Fock levels, simulated with CUDA-Q Dynamics \cite{kim2023cuda}.
The Hamiltonian is
\begin{equation*}
 H = \omega_c a^{\dagger}a + \omega_q \sigma_+\sigma_-
   + g(\sigma_-a^{\dagger}+\sigma_+a),
\end{equation*}
with $\omega_c=\omega_q=2\pi$ and $g=\pi$. Photon loss uses $C=\sqrt{\gamma}\,a$ with $\gamma=0.05$. The initial state is $\rho_0=|g,1\rangle\langle g,1|$, and the target observable is qubit excitation expectation probability $\langle\sigma_+\sigma_-\rangle(t)$ for $t\in[0,50]$.

\paragraph{Dispersive transmon--resonator dynamics}
The second task is a closed dispersive transmon--resonator model with a two-level transmon and a resonator truncated to 20 Fock levels, also simulated with CUDA-Q Dynamics. The Hamiltonian is
\begin{equation*}
 H
 =
 \frac{1}{2}\omega'_{01}\sigma_z
 +
 (\omega'_r+\chi\sigma_z)a^{\dagger}a .
\end{equation*}
We use $\omega_{01}=3.0\cdot 2\pi$ GHz, $\omega_r=2.0\cdot 2\pi$ GHz, $\chi=0.025\cdot 2\pi$ GHz, $\omega'_{01}=\omega_{01}+\chi$, and $\omega'_r=\omega_r$. The initial state is $(|0\rangle+|1\rangle)/\sqrt{2}$ for the transmon and $|\alpha=2.0\rangle$ for the resonator. The target is the expectation value of the resonator's position quadrature operator $\langle \hat{x}\rangle(t)$ for $t\in[0,25]$ ns. Both CUDA-Q Dynamics trajectories contain 3000 equally spaced time steps.

\subsection{Single-step prediction}
For input sequence length $N$, each input is $\mathbf{x}_{t,N}=[x_{t-N},\ldots,x_{t-1}]$ and the target is $x_t$. The model processes the $N$ observations sequentially, updates the fast state at each internal step, and predicts after the final observation. We evaluate $N\in\{4,8,16,32,64\}$ to test how each update rule retains information as sequence length grows.

\subsection{Direct multi-step forecasting}
For direct forecasting, the input length is fixed to $N=64$ and the model outputs the entire horizon in one forward pass:
$\mathbf{x}_{t,64}=[x_{t-64},\ldots,x_{t-1}]$ and $\mathbf{y}_{t,H}=[x_t,x_{t+1},\ldots,x_{t+H-1}]$, with $H\in\{4,8,16\}$. The protocol avoids recursive feedback of predictions and isolates whether the fast state contains enough history for longer-horizon prediction.

\subsection{Training and evaluation}
We use a GPU-efficient \texttt{FlashQKAN} implementation in PyTorch \cite{paszke2019pytorchimperativestylehighperformance}, accelerated with \texttt{CuTe DSL} \cite{cecka2026cutelayoutrepresentationalgebra} for fused operators and block tiling on CUDA devices, building on the open-source QKAN repository \cite{jiang2025qkan_github}.
For each dataset, architecture, update rule, sequence length, and horizon, we train with five independent random seeds for 100 epochs using Adam \cite{kingma2014adam} with learning rate $10^{-3}$ and batch size $4$. The primary metric is MSE. For multi-step forecasting, MSE over the horizon is calculated as
$ \mathrm{MSE}
  =
  \frac{1}{H}
  \sum_{h=0}^{H-1}
  \left(\hat{x}_{t+h}-x_{t+h}\right)^2,$
then averaged across samples and seeds. Relative improvement over the scalar-gated baseline is
\begin{equation}
  \Delta_{\mathrm{rel}}
  =
  \frac{M_{\mathrm{gated}}-M_{\mathrm{self-modulated}}}
     {M_{\mathrm{gated}}+10^{-12}} .
  \label{eq:relative_improvement}
\end{equation}
Positive values indicate lower MSE than the corresponding scalar-gated model.

\begin{table}[t]
\centering
\scriptsize
\setlength{\tabcolsep}{2.6pt}
\renewcommand{\arraystretch}{0.94}
\caption{Best models at $N=16$ for each dataset. Final test MSE is reported as mean (standard deviation) over five seeds.}
\label{tab:best-arm-window16}
\resizebox{\columnwidth}{!}{
\begin{tabular}{llllc}
\toprule
Dataset & Selected arm & Family & Rule & Epoch-100 test MSE \\
\midrule
Bessel & CMG QKAN-QKANFWP & QKAN-QKANFWP & CMG & \(4.94{\scriptstyle(6.7)}\times10^{-6}\) \\
Damped SHM & CMG FWP & FWP & CMG & \(1.42{\scriptstyle(1.2)}\times10^{-5}\) \\
Narma5 & G-QKANFWP~\cite{peng2026gatedqkanfwp} & QKANFWP & Gated & \(5.21{\scriptstyle(3.9)}\times10^{-6}\) \\
Narma10 & Only-new QKANFWP & QKANFWP & Only New & \(2.55{\scriptstyle(2.5)}\times10^{-6}\) \\
DQC & CMG QKANFWP & QKANFWP & CMG & \(6.94{\scriptstyle(3.5)}\times10^{-6}\) \\
Jaynes-Cummings & CMG QKAN-QKANFWP & QKAN-QKANFWP & CMG & \(3.18{\scriptstyle(1.6)}\times10^{-6}\) \\
transmon-resonator & CMG QKAN-QKANFWP & QKAN-QKANFWP & CMG & \(2.65{\scriptstyle(1.9)}\times10^{-6}\) \\
\bottomrule
\end{tabular}
}
\vspace{-10pt}
\end{table}

\begin{table}[t]
\centering
\scriptsize
\caption{Final test MSE of the selected models on the smooth-dynamics benchmarks for $N=\{4,8,32,64\}$. Values are reported as mean (standard deviation) over five seeds. Best/second-best results are shown in bold/underlined.}
\label{tab:selected-winners-smooth}
\resizebox{\columnwidth}{!}{
\begin{tabular}{lcccc}
\toprule
Model & $N=4$ & $N=8$ & $N=32$ & $N=64$ \\
\midrule
\multicolumn{5}{l}{\textit{Bessel}}\\
CMG QKAN-QKANFWP & \(\mathbf{6.24{\scriptstyle(3.3)}\times10^{-6}}\) & \(\underline{1.86{\scriptstyle(3.2)}\times10^{-5}}\) & \(\underline{1.92{\scriptstyle(2.1)}\times10^{-5}}\) & \(\underline{8.37{\scriptstyle(5.3)}\times10^{-6}}\) \\
CMG FWP & \(3.23{\scriptstyle(2.6)}\times10^{-5}\) & \(1.86{\scriptstyle(1.9)}\times10^{-5}\) & \(6.07{\scriptstyle(11.7)}\times10^{-4}\) & \(2.04{\scriptstyle(3.1)}\times10^{-5}\) \\
G-QKANFWP~\cite{peng2026gatedqkanfwp} & \(8.15{\scriptstyle(15.3)}\times10^{-4}\) & \(4.08{\scriptstyle(5.0)}\times10^{-5}\) & \(1.23{\scriptstyle(1.5)}\times10^{-3}\) & \(1.47{\scriptstyle(1.8)}\times10^{-3}\) \\
Only-new QKANFWP & \(\underline{1.49{\scriptstyle(1.3)}\times10^{-5}}\) & \(6.38{\scriptstyle(11.4)}\times10^{-5}\) & \(3.53{\scriptstyle(0.3)}\times10^{-3}\) & \(3.45{\scriptstyle(0.6)}\times10^{-3}\) \\
CMG QKANFWP & \(1.67{\scriptstyle(1.2)}\times10^{-5}\) & \(\mathbf{8.73{\scriptstyle(4.6)}\times10^{-6}}\) & \(\mathbf{2.90{\scriptstyle(2.2)}\times10^{-6}}\) & \(\mathbf{6.37{\scriptstyle(11.1)}\times10^{-6}}\) \\
\midrule
\multicolumn{5}{l}{\textit{Damped SHM}}\\
CMG QKAN-QKANFWP & \(\underline{1.16{\scriptstyle(1.3)}\times10^{-4}}\) & \(\underline{1.43{\scriptstyle(1.4)}\times10^{-5}}\) & \(\mathbf{3.60{\scriptstyle(1.4)}\times10^{-6}}\) & \(\underline{1.87{\scriptstyle(2.7)}\times10^{-5}}\) \\
CMG FWP & \(4.41{\scriptstyle(6.5)}\times10^{-4}\) & \(3.24{\scriptstyle(2.4)}\times10^{-5}\) & \(2.85{\scriptstyle(2.9)}\times10^{-5}\) & \(6.00{\scriptstyle(7.6)}\times10^{-5}\) \\
G-QKANFWP~\cite{peng2026gatedqkanfwp} & \(1.14{\scriptstyle(1.2)}\times10^{-3}\) & \(1.14{\scriptstyle(1.2)}\times10^{-3}\) & \(1.03{\scriptstyle(1.2)}\times10^{-3}\) & \(1.21{\scriptstyle(1.4)}\times10^{-3}\) \\
Only-new QKANFWP & \(1.18{\scriptstyle(1.4)}\times10^{-4}\) & \(3.87{\scriptstyle(4.2)}\times10^{-5}\) & \(2.08{\scriptstyle(1.9)}\times10^{-4}\) & \(9.04{\scriptstyle(6.9)}\times10^{-5}\) \\
CMG QKANFWP & \(\mathbf{5.97{\scriptstyle(2.5)}\times10^{-5}}\) & \(\mathbf{1.17{\scriptstyle(0.7)}\times10^{-5}}\) & \(\underline{2.51{\scriptstyle(3.7)}\times10^{-5}}\) & \(\mathbf{1.00{\scriptstyle(1.1)}\times10^{-5}}\) \\
\bottomrule
\end{tabular}
}
\vspace{-15pt}
\end{table}

\begin{table}[t]
\centering
\scriptsize
\caption{Final test MSE of the selected models on the Narma benchmarks for $N=\{4,8,32,64\}$. Values are reported as mean (standard deviation) over five seeds. Best/second-best results are shown in bold/underlined.}
\label{tab:selected-winners-narma}
\resizebox{\columnwidth}{!}{
\begin{tabular}{lcccc}
\toprule
Model & $N=4$ & $N=8$ & $N=32$ & $N=64$ \\
\midrule
\multicolumn{5}{l}{\textit{Narma-5}}\\
CMG QKAN-QKANFWP & \(3.15{\scriptstyle(1.8)}\times10^{-5}\) & \(1.90{\scriptstyle(1.9)}\times10^{-5}\) & \(\underline{1.34{\scriptstyle(1.1)}\times10^{-5}}\) & \(1.57{\scriptstyle(1.2)}\times10^{-5}\) \\
CMG FWP & \(2.45{\scriptstyle(1.1)}\times10^{-5}\) & \(1.32{\scriptstyle(1.3)}\times10^{-5}\) & \(1.59{\scriptstyle(1.6)}\times10^{-5}\) & \(1.48{\scriptstyle(0.6)}\times10^{-5}\) \\
G-QKANFWP~\cite{peng2026gatedqkanfwp} & \(\mathbf{1.70{\scriptstyle(0.9)}\times10^{-5}}\) & \(1.05{\scriptstyle(1.2)}\times10^{-5}\) & \(1.99{\scriptstyle(2.2)}\times10^{-5}\) & \(\underline{1.38{\scriptstyle(0.7)}\times10^{-5}}\) \\
Only-new QKANFWP & \(\underline{2.04{\scriptstyle(0.7)}\times10^{-5}}\) & \(\mathbf{4.91{\scriptstyle(6.7)}\times10^{-6}}\) & \(5.90{\scriptstyle(3.1)}\times10^{-5}\) & \(4.72{\scriptstyle(1.5)}\times10^{-5}\) \\
CMG QKANFWP & \(2.62{\scriptstyle(0.8)}\times10^{-5}\) & \(\underline{8.94{\scriptstyle(10.0)}\times10^{-6}}\) & \(\mathbf{7.13{\scriptstyle(3.6)}\times10^{-6}}\) & \(\mathbf{9.58{\scriptstyle(5.0)}\times10^{-6}}\) \\
\midrule
\multicolumn{5}{l}{\textit{Narma-10}}\\
CMG QKAN-QKANFWP & \(1.16{\scriptstyle(0.5)}\times10^{-4}\) & \(4.60{\scriptstyle(3.1)}\times10^{-5}\) & \(2.09{\scriptstyle(3.0)}\times10^{-5}\) & \(2.84{\scriptstyle(3.4)}\times10^{-5}\) \\
CMG FWP & \(\underline{1.14{\scriptstyle(0.3)}\times10^{-4}}\) & \(\mathbf{2.25{\scriptstyle(0.5)}\times10^{-5}}\) & \(2.47{\scriptstyle(2.1)}\times10^{-5}\) & \(1.57{\scriptstyle(0.9)}\times10^{-5}\) \\
G-QKANFWP~\cite{peng2026gatedqkanfwp} & \(\mathbf{8.36{\scriptstyle(2.3)}\times10^{-5}}\) & \(4.79{\scriptstyle(1.9)}\times10^{-5}\) & \(\mathbf{1.33{\scriptstyle(0.7)}\times10^{-5}}\) & \(\underline{1.50{\scriptstyle(0.5)}\times10^{-5}}\) \\
Only-new QKANFWP & \(1.17{\scriptstyle(0.4)}\times10^{-4}\) & \(4.04{\scriptstyle(2.1)}\times10^{-5}\) & \(1.22{\scriptstyle(0.6)}\times10^{-4}\) & \(8.39{\scriptstyle(1.5)}\times10^{-5}\) \\
CMG QKANFWP & \(1.37{\scriptstyle(0.7)}\times10^{-4}\) & \(\underline{2.68{\scriptstyle(0.9)}\times10^{-5}}\) & \(\underline{1.48{\scriptstyle(1.1)}\times10^{-5}}\) & \(\mathbf{1.36{\scriptstyle(0.5)}\times10^{-5}}\) \\
\bottomrule
\end{tabular}
}
\vspace{-10pt}
\end{table}

\begin{table}[t]
\centering
\scriptsize
\caption{Final test MSE of the selected models on the quantum-dynamics benchmarks for $N=\{4,8,32,64\}$. Values are reported as mean (standard deviation) over five seeds. Best/second-best results are shown in bold/underlined.}
\label{tab:selected-winners-quantum}
\resizebox{\columnwidth}{!}{
\begin{tabular}{lcccc}
\toprule
Model & $N=4$ & $N=8$ & $N=32$ & $N=64$ \\
\midrule
\multicolumn{5}{l}{\textit{DQC}}\\
CMG QKAN-QKANFWP & \(\mathbf{4.02{\scriptstyle(2.8)}\times10^{-6}}\) & \(\mathbf{3.52{\scriptstyle(1.9)}\times10^{-6}}\) & \(\mathbf{7.82{\scriptstyle(4.0)}\times10^{-6}}\) & \(\mathbf{4.77{\scriptstyle(1.2)}\times10^{-6}}\) \\
CMG FWP & \(2.44{\scriptstyle(2.8)}\times10^{-5}\) & \(1.07{\scriptstyle(1.3)}\times10^{-5}\) & \(\underline{1.16{\scriptstyle(0.7)}\times10^{-5}}\) & \(1.96{\scriptstyle(1.9)}\times10^{-5}\) \\
G-QKANFWP~\cite{peng2026gatedqkanfwp} & \(6.42{\scriptstyle(6.9)}\times10^{-5}\) & \(1.21{\scriptstyle(0.1)}\times10^{-4}\) & \(1.19{\scriptstyle(0.2)}\times10^{-4}\) & \(1.47{\scriptstyle(0.2)}\times10^{-4}\) \\
Only-new QKANFWP & \(1.24{\scriptstyle(0.7)}\times10^{-5}\) & \(1.10{\scriptstyle(1.0)}\times10^{-5}\) & \(8.81{\scriptstyle(9.0)}\times10^{-5}\) & \(3.33{\scriptstyle(1.9)}\times10^{-5}\) \\
CMG QKANFWP & \(\underline{4.04{\scriptstyle(1.8)}\times10^{-6}}\) & \(\underline{4.57{\scriptstyle(3.4)}\times10^{-6}}\) & \(1.19{\scriptstyle(1.1)}\times10^{-5}\) & \(\underline{1.44{\scriptstyle(1.0)}\times10^{-5}}\) \\
\midrule
\multicolumn{5}{l}{\textit{Jaynes-Cummings}}\\
CMG QKAN-QKANFWP & \(5.42{\scriptstyle(2.8)}\times10^{-5}\) & \(\underline{3.17{\scriptstyle(1.5)}\times10^{-5}}\) & \(\underline{1.04{\scriptstyle(1.3)}\times10^{-5}}\) & \(\underline{1.47{\scriptstyle(1.6)}\times10^{-5}}\) \\
CMG FWP & \(4.31{\scriptstyle(2.9)}\times10^{-5}\) & \(\mathbf{1.66{\scriptstyle(0.9)}\times10^{-5}}\) & \(5.21{\scriptstyle(3.3)}\times10^{-5}\) & \(1.04{\scriptstyle(1.1)}\times10^{-4}\) \\
G-QKANFWP~\cite{peng2026gatedqkanfwp} & \(4.27{\scriptstyle(3.1)}\times10^{-4}\) & \(7.02{\scriptstyle(1.0)}\times10^{-4}\) & \(7.30{\scriptstyle(0.9)}\times10^{-4}\) & \(7.98{\scriptstyle(1.4)}\times10^{-4}\) \\
Only-new QKANFWP & \(\underline{3.56{\scriptstyle(3.3)}\times10^{-5}}\) & \(1.40{\scriptstyle(1.2)}\times10^{-4}\) & \(1.26{\scriptstyle(0.8)}\times10^{-3}\) & \(1.53{\scriptstyle(0.8)}\times10^{-3}\) \\
CMG QKANFWP & \(\mathbf{3.47{\scriptstyle(3.4)}\times10^{-5}}\) & \(3.98{\scriptstyle(3.1)}\times10^{-5}\) & \(\mathbf{9.10{\scriptstyle(10.8)}\times10^{-6}}\) & \(\mathbf{7.04{\scriptstyle(7.0)}\times10^{-6}}\) \\
\midrule
\multicolumn{5}{l}{\textit{transmon-resonator}}\\
CMG QKAN-QKANFWP & \(\mathbf{1.08{\scriptstyle(0.9)}\times10^{-5}}\) & \(\underline{1.10{\scriptstyle(1.5)}\times10^{-5}}\) & \(\underline{8.89{\scriptstyle(11.1)}\times10^{-6}}\) & \(\underline{6.90{\scriptstyle(6.3)}\times10^{-6}}\) \\
CMG FWP & \(2.35{\scriptstyle(1.9)}\times10^{-5}\) & \(4.97{\scriptstyle(7.3)}\times10^{-5}\) & \(1.68{\scriptstyle(1.5)}\times10^{-5}\) & \(1.78{\scriptstyle(1.5)}\times10^{-5}\) \\
G-QKANFWP & \(2.75{\scriptstyle(2.2)}\times10^{-3}\) & \(4.67{\scriptstyle(0.2)}\times10^{-3}\) & \(3.74{\scriptstyle(1.9)}\times10^{-3}\) & \(3.77{\scriptstyle(1.9)}\times10^{-3}\) \\
Only-new QKANFWP & \(\underline{1.56{\scriptstyle(0.5)}\times10^{-5}}\) & \(2.14{\scriptstyle(1.2)}\times10^{-5}\) & \(3.80{\scriptstyle(3.7)}\times10^{-5}\) & \(4.12{\scriptstyle(1.6)}\times10^{-3}\) \\
CMG QKANFWP & \(4.64{\scriptstyle(6.8)}\times10^{-5}\) & \(\mathbf{2.49{\scriptstyle(1.4)}\times10^{-6}}\) & \(\mathbf{7.09{\scriptstyle(3.4)}\times10^{-6}}\) & \(\mathbf{6.40{\scriptstyle(4.6)}\times10^{-6}}\) \\
\bottomrule
\end{tabular}
}
\vspace{-10pt}
\end{table}

\begin{table}[t]
\centering
\scriptsize
\setlength{\tabcolsep}{3.2pt}
\renewcommand{\arraystretch}{0.94}
\caption{Parameter counts and area under the test-MSE learning curve (AULC) for single-step forecasting at $N=64$. AULC values are averaged over the Jaynes–Cummings and transmon–resonator datasets and five seeds. Parameter ratios are relative to the corresponding gated model. Lower AULC is better.}
\label{tab:complexity-speed}
\resizebox{\columnwidth}{!}{
\begin{tabular}{llcccc}
\toprule
Model & Update & Params & Params/Gated & Test-MSE AULC $\downarrow$ \\
\midrule
FWP & Only-old & 209 & 152.6\% & \(2.41\times10^{-4}\) \\
 & Full & 299 & 218.2\% & \(2.86\times10^{-4}\) \\
 & CMG & 209 & 152.6\% & \(9.17\times10^{-5}\) \\
QKANFWP & Only-old & 210 & 162.8\% & \(9.48\times10^{-5}\) \\
 & Full & 300 & 232.6\% & \(1.02\times10^{-4}\) \\
 & CMG & 210 & 162.8\% & \(\underline{4.45\times10^{-5}}\)\\
QKAN-QKANFWP & Only-old & 234 & 139.3\% & \(1.20\times10^{-4}\) \\
 & Full & 306 & 182.1\% & \(1.37\times10^{-4}\) \\
 & CMG & 234 & 139.3\% & \(\mathbf{2.99\times10^{-5}}\) \\
\bottomrule
\end{tabular}
}
\vspace{-10pt}
\end{table}

\section{Experimental Results and Analysis}
\label{sec:results}

\subsection{Single-step prediction}

\begin{figure*}[!t]
  \centering
  \includegraphics[width=\textwidth,trim={0 15pt 0 10pt}]{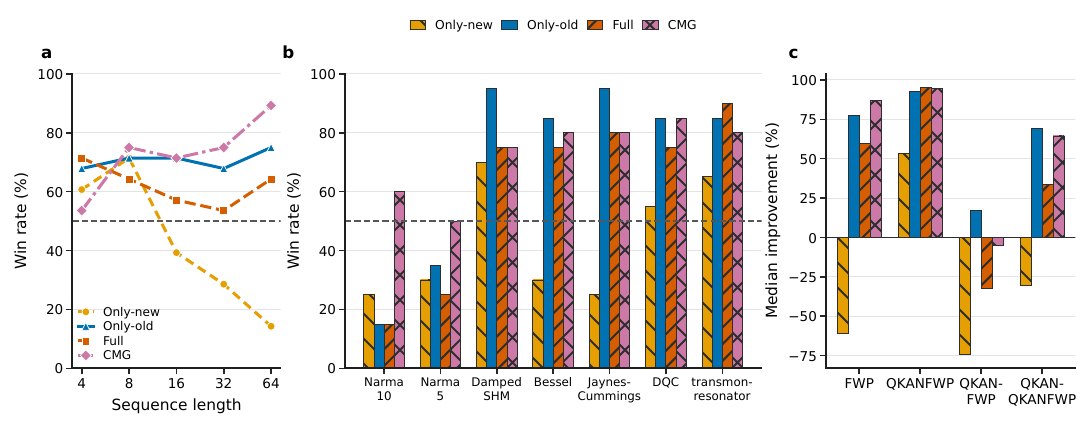}
  \caption{Performance of the self-modulating update rules relative to their paired gated baselines. (a) Win rate by sequence length, aggregated across datasets and model families. (b) Win rate by dataset, aggregated across sequence lengths and model families. (c) Median relative MSE improvement [~\cref{eq:relative_improvement}] by model family, aggregated across datasets and sequence lengths. Dashed lines in (a) and (b) mark a 50\% win rate; the horizontal line in (c) marks zero improvement.}
  \label{fig:summary}
\vspace{-10pt}
\end{figure*}

\begin{figure*}[t]
  \centering
  \includegraphics[width=\textwidth]{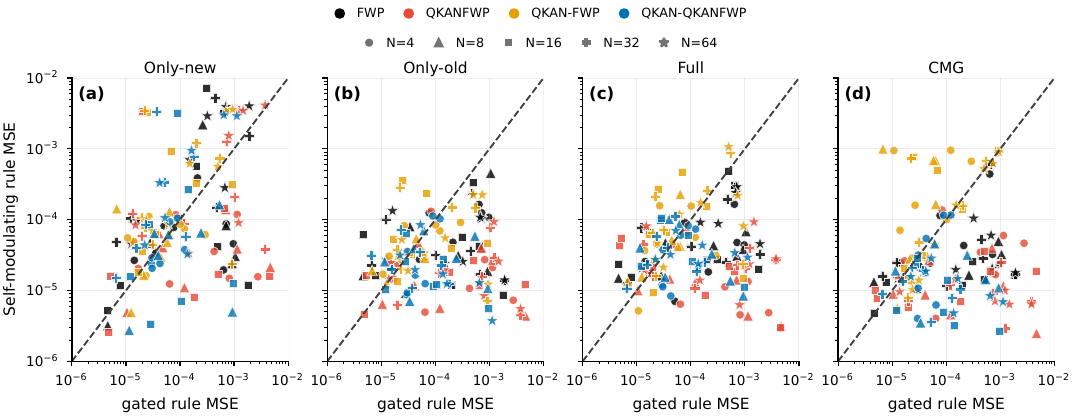}
  \caption{Final test MSE of each self-modulating rule versus its paired gated baseline across all datasets, sequence lengths, and model families: (a) Only-new, (b) Only-old, (c) Full, and (d) CMG. The diagonal indicates equal MSE; points below it favor self-modulation. Colors denote model families, and markers denote sequence lengths.}
  \label{fig:scatter}
  \vspace{-15pt}
\end{figure*}

We first evaluate all update rules across the four model families and sequence lengths $N \in \{4,8,16,32,64\}$. \Cref{fig:summary} establishes that CMG provides the most consistent improvement over the scalar-gated baseline.
The paired scatter plots in \cref{fig:scatter} support the same conclusion. 
CMG produces the cleanest shift below the diagonal, especially for QKANFWP and QKAN-QKANFWP. 
In contrast, Only-new has many failures above the diagonal, Only-old mainly confirms the value of memory-side modulation, and Full shows that adding both old- and new-side modulation is not always necessary. 
QKAN-FWP is the main exception. 
Its scalar-gated form is already competitive, and the self-modulation variants provide only marginal or inconsistent gains. 
Thus, CMG's benefit is best interpreted as stable coordinate-wise old/new balancing that is most useful when the quantum-inspired structure is used directly in the fast programmer.

Following prior experimental conventions~\cite{peng2026gatedqkanfwp}, we isolate the top-performing model at $N=16$ for each dataset and re-evaluate these selected arms across the remaining sequence lengths as reported in~\cref{tab:best-arm-window16,tab:selected-winners-smooth,tab:selected-winners-narma,tab:selected-winners-quantum}. The cross-sequence results strongly favor the QKAN-based backbones over classical FWP. 
Across the 28 configurations, CMG variants achieve the best MSE in 24 cases. 
Specifically, CMG QKAN-QKANFWP and CMG QKANFWP decisively outperform classical FWP across most tasks. 
While the NARMA datasets exhibit slight heterogeneity at shorter sequences, CMG QKANFWP still achieves the best MSE at the longest sequence length ($N=64$), confirming that CMG becomes increasingly critical as the effective context window grows.

\begin{figure}[t]
  \centering
  \includegraphics[width=\columnwidth]{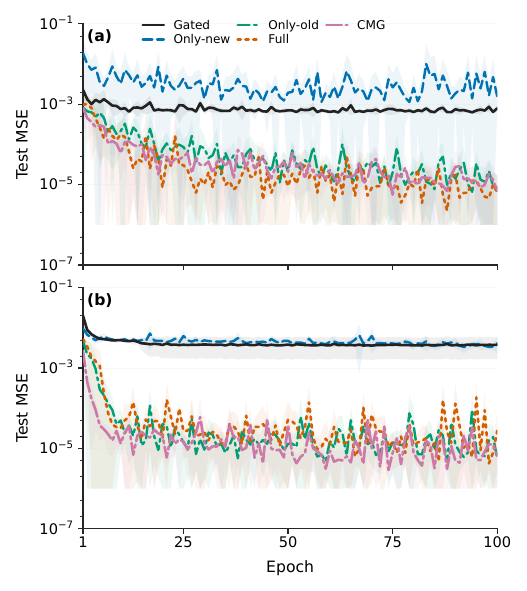}
  \vspace{-20pt}
  \caption{Test MSE versus training epoch for QKANFWP at $N=64$: (a) Jaynes–Cummings and (b) transmon–resonator. Curves show the mean over five seeds, and shaded bands show $\pm 1$ standard deviation.}
  \label{fig:learning_curve}
\vspace{-10pt}
\end{figure}

Finally, we analyze learning dynamics at the longest sequence length ($N=64$) on the two CUDA-Q dynamics datasets. 
\Cref{fig:learning_curve} shows that CMG's advantage is not merely a final-epoch artifact. On Jaynes--Cummings, CMG stays among the lowest-error rules after early training. On transmon--resonator, it reaches the low-error regime quickly and remains close to Only-old and Full through epoch 100. 
In contrast, scalar gating and Only-new remain in a higher-error regime, especially on transmon--resonator.
\Cref{tab:complexity-speed} further contextualizes this by comparing parameter counts and the Area Under the Learning Curve (AULC)~\cite{viering2022shape,mazzoni2004active} for the self-modulating variants. CMG matches the parameter efficiency of Only-old while requiring fewer parameters than Full self-modulation. 
In these benchmarks, CMG QKANFWP and CMG QKAN-QKANFWP yield the lowest test-MSE AULC, indicating lower average error throughout training. 
Based on the single-step results, we select the two top-performing models, QKANFWP and QKAN-QKANFWP for the subsequent direct multi-step prediction tasks.

\subsection{Direct multi-step prediction}
We next evaluate direct multi-step forecasting on the two CUDA-Q Dynamics datasets with $N=64$ and $H\in\{4,8,16\}$. We focus on QKANFWP and QKAN-QKANFWP because the single-step study identifies them as the most robust HQKAN-based families.

\begin{figure}[t]
  \centering
  \includegraphics[width=\columnwidth]{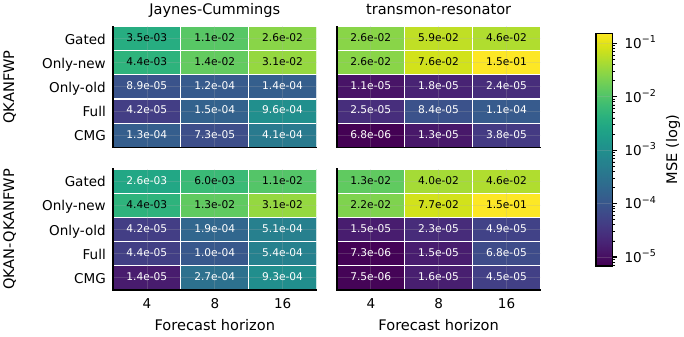}
  \caption{Mean test MSE for direct multi-step forecasting with QKANFWP and QKAN-QKANFWP at $N=64$, separated by dataset, forecast horizon, and update rule. Values are averaged over five seeds; colors encode MSE on a logarithmic scale. CMG uses one low-rank-generated gate matrix, whereas Full uses separate old- and new-modulation matrices.}
  \label{fig:mse_multi_step}
  \vspace{-10pt}
\end{figure}

\begin{figure}[t]
  \centering
  \includegraphics[width=\columnwidth]{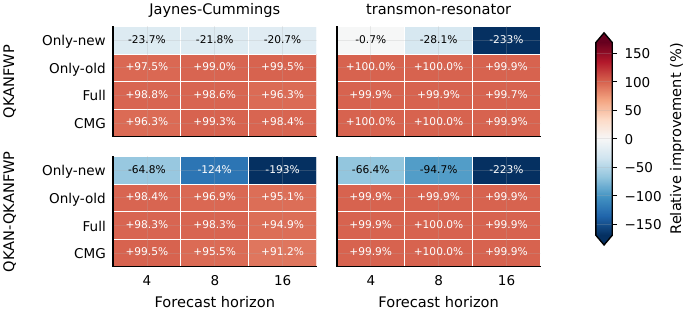}
  \caption{Relative MSE improvement over the paired gated baseline [~\cref{eq:relative_improvement}] for direct multi-step forecasting at $N=64$. Rows group update rules within QKANFWP and QKAN-QKANFWP; columns group forecast horizons within the Jaynes–Cummings and transmon–resonator datasets. Values are computed from mean MSE over five seeds. Red denotes improvement, whereas blue denotes degradation.}
  \label{fig:improve_multi_step}
\vspace{-10pt}
\end{figure}

The heatmaps in \cref{fig:mse_multi_step,fig:improve_multi_step} show that memory-side element-wise updates remain strong in long-horizon forecasting. For CMG, every QKANFWP and QKAN-QKANFWP cell is no worse than $9.3\times10^{-4}$, placing the direct expectation-value forecasts at the $10^{-4}$ scale rather than the $10^{-2}$ scale of scalar gating. Only-old, Full, and CMG reduce MSE by orders of magnitude over scalar gating on both quantum systems, whereas Only-new is worse than the gated baseline in every multi-step setting. CMG is the best rule in five of the twelve dataset-horizon cells: QKANFWP on Jaynes--Cummings at $H=8$, QKANFWP on transmon--resonator at $H=4,8$, QKAN-QKANFWP on Jaynes--Cummings at $H=4$, and QKAN-QKANFWP on transmon--resonator at $H=16$. Full is best in four cells, and Only-old in three. Thus CMG reaches the same competitive regime with the same modulation-head size as Only-old and fewer heads than Full. 
The relative-improvement heatmap clarifies the Only-new failure mode. For QKAN-QKANFWP, Only-new worsens Jaynes--Cummings by $64.8\%$, $124\%$, and $193\%$ for $H=4,8,16$, and worsens transmon--resonator by $66.4\%$, $94.7\%$, and $223\%$. In contrast, CMG achieves at least $91.2\%$ relative improvement across all multi-step cells and reaches $99.9\%$--$100\%$ improvement on transmon--resonator. These results indicate that direct forecasting benefits from stable retention or complementary retention/write gating, rather than write-only modulation.

\begin{figure}[t]
  \centering
  \includegraphics[width=\columnwidth]{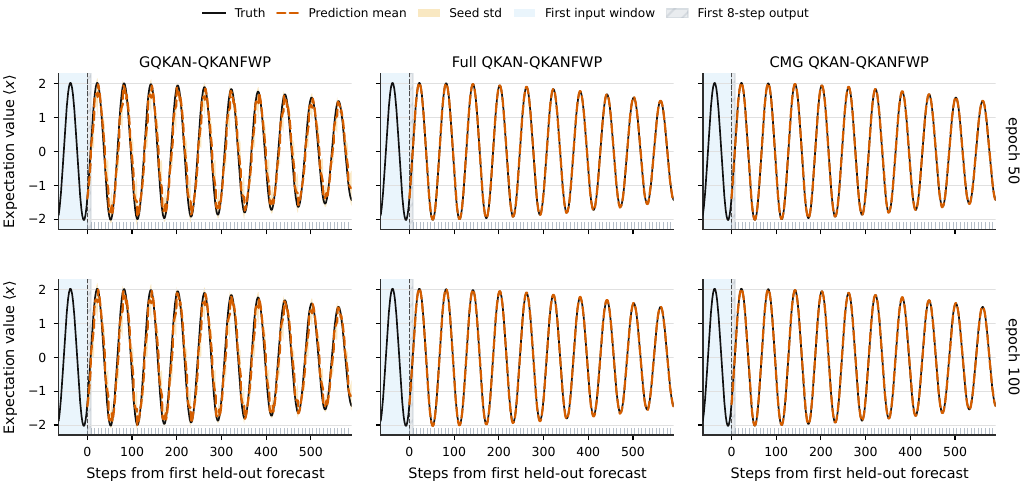}
  \caption{Direct multi-step forecasts on the transmon–resonator test set for QKAN-QKANFWP with $N=64$ and $H=8$. Columns compare scalar-gated, Full, and CMG updates; rows show model checkpoints after 50 and 100 epochs. Curves show the mean prediction over five seeds with $\pm1$ standard-deviation bands and the ground truth. Shading identifies the first input window and first eight-step output block.}
  \label{fig:fit}
\vspace{-10pt}
\end{figure}

The qualitative forecasts in \cref{fig:fit} match the quantitative results. Full and CMG QKAN-QKANFWP track the held-out oscillatory trajectory with accurate phase and amplitude over non-overlapping forecast blocks. The gated model captures the broad oscillation but shows larger phase/amplitude mismatch and wider seed variation. CMG remains visually comparable to Full despite using only one matrix-gating head.

\section{Conclusion}
We presented Self-Modulating QKAN-FWP, a framework that equips quantum-inspired fast-weight programmers with coordinate-wise memory control, and CMG as its central update rule. CMG generalizes the scalar gate of gated QKAN-FWPs into a single low-rank-generated matrix gate whose complement writes the new proposal. It therefore preserves the two properties that make scalar gating attractive---coordinate-wise bounded-convex stability and affine parallel-prefix-scan compatibility---while relaxing the constraint that one coefficient govern every fast-state coordinate, at the modulation-head cost of Only-old and below that of Full self-modulation.

Across seven single-step benchmarks and five sequence lengths, CMG is the most consistent update rule for QKANFWP and QKAN-QKANFWP, and in direct multi-step forecasting of Jaynes--Cummings and transmon--resonator dynamics it holds expectation-value error at the $10^{-4}$ scale or below across $H\in\{4,8,16\}$, improving on scalar gating by at least $91.2\%$. Together with the failure of write-only modulation at long horizons, these results indicate that the decisive ingredient is not additional modulation capacity but a stable, complementary balance between retained memory and new writes.

The present formulation targets univariate temporal memory and does not explicitly represent spatial structure among variables. Extending complementary matrix gating to structured spatio-temporal fast states is a natural next step toward multivariate physical dynamics and quantum-control forecasting.

\ifarxiv
\section*{Acknowledgment}
K.-C. Peng, J.-C. Jiang, Y.-C. Hsu and C.-H. Lin thank the National Center for High-Performance Computing (NCHC), National Institutes of Applied Research (NIAR), Taiwan, for providing computational and storage resources supported by the National Science and Technology Council (NSTC), Taiwan, under Grants No. NSTC 114-2119-M-007-013.
H.-S. Goan acknowledges support from the NSTC, Taiwan, under Grants No. NSTC 113-2112-M-002-022-MY3, No. NSTC 113-2119-M-002-021, No. NSTC 114-2119-M-002-018, No. NSTC 114-2119-M-002-017-MY3, and from the National Taiwan University under Grants No. NTU-CC-115L8937, No. NTU-CC-115L893704 and No. NTU-CC-115L8512. H.-S. Goan is also grateful for the support of the Center for Advanced Computing and Imaging in Biomedicine through the Featured Areas Research Center Program within the framework of the Higher Education Sprout Project by the Ministry of Education, Taiwan, the support of Taiwan Semiconductor Research Institute through the Joint Developed Project and the support from the Physics Division, National Center for Theoretical Sciences, Taiwan. E.-J. Kuo acknowledges financial support from the NSTC of Taiwan under Grant No.~NSTC~114-2112-M-A49-036-MY3.
\fi

\bibliographystyle{IEEEtran}
\bibliography{ref}

\end{document}